\documentclass[11pt,a4paper]{article}
\usepackage{jcappub}

\begin{document}
	
\title{How Gaussian can the Sky be? Primordial Non-Gaussianity from Quantum Information}

\author[a]{C\'esar G\'omez,} 
\author[b,c]{Raul Jimenez} 

\affiliation[a]{Instituto de F\'{i}sica Te\'orica UAM-CSIC, Universidad Aut\'onoma de Madrid, Cantoblanco, 28049 Madrid, Spain.}
\affiliation[b]{ICC, University of Barcelona, Marti i Franques 1, 08028 Barcelona, Spain}
\affiliation[c]{ICREA, Pg. Lluis Companys 23, Barcelona, E-08010, Spain.}

\emailAdd{bfkl08@gmail.com}
\emailAdd{raul.jimenez@icc.ub.edu}

\abstract{
Using the quantum information picture to describe the early universe as a time dependent quantum density matrix, with time playing the role of a stochastic variable, we compute the non-gaussian features in the distribution of primordial fluctuations.  We use a quasi de Sitter model to compute the corresponding quantum Fisher information function as the second derivative of the relative entanglement entropy for the density matrix at two different times. We define the curvature fluctuations in terms of the time quantum estimator. Using standard quantum estimation theory we compute the non-gaussian features in the  statistical distribution of primordial fluctuations. Our approach is model independent and only relies on the existence of a quasi de Sitter phase. We show that there are primordial non-gaussianities, both in the form of squeezed and equilateral shapes. The squeezed limit gives a value of $f_{\rm NL} \sim n_s-1$. In the equilateral limit we find that $f_{\rm NL} \sim 0.03$. The equilateral non-gaussianity is due to the non-linearity of Einstein's equation. On the other hand, the squeezed one is due to the quantum nature of clock synchronization and thus real and cannot be gauged away as a global curvature. We identify a new effect: {\it clock bias} which is a pure quantum effect and introduces a bias in the spectral tilt and running of the power spectrum of order $\sim 10^{-4}$, which could be potentially measurable and yield precious information on the quantum nature of the early Universe.}

\maketitle

\section{Introduction}
In the current cosmological model ($\Lambda$CDM see e.g. Ref.~\cite{Planck18}) primordial fluctuations are generated as quantum fluctuations during an early inflationary period.  These fluctuations happen for both the inflaton field and the space-time metric. These fluctuations later classicalise as they re-enter the horizon~\cite{Planck18,Starobinsky,Guth,Linde,Mukhanov,Hawking,Albrecht,GuthPi,WMAP03}. The statistical distribution of these fluctuations is nearly gaussian and indeed current observational constraints indicate this to be the case~\cite{Planck18NG}. However, even for the simplest inflationary models, some small deviations from non-gaussianity are expected~\cite{BartoloReview}. The key question is: what is the physics that these  non-gaussianities represent?  and, can they be actually measured?~\cite{NGgauge}.

The current observational constraints on non-gaussianity come from measurements of the cosmic microwave background temperature and E-mode polarization higher-order correlation functions using the Planck18 mission~\cite{Planck18NG}; it already constraints non-gaussianity to be very small. The current limits, as there is no detection, for the different shapes are: $f_{\rm NL}^{\rm local} = -0.9 \pm 5.1$; $f_{\rm NL}^{\rm equil} = -26 \pm 47$; and $f_{\rm NL}^{\rm ortho} = - 38 \pm 24$ (at the 68\% confidence level). Where $f_{\rm NL}$ is the first coefficient in the Taylor expansion of the gravitational Bardeen potential at linear order: $\Phi = \Phi_{\rm L} + f_{\rm NL} (\Phi_{\rm L}^2 - <\Phi_{\rm L}^2>)$. Recall that $f_{\rm NL}$ multiplies $\Phi_L^2 \sim 10^{-10}$, thus the level of non-gaussianity is already an extremely tiny correction to the Gaussian distribution; a correction of more than 5 orders of magnitude. The number of inflationary models already restricted by the above constraints on non-gaussianity is vast~\cite{Planck18NG}. 

Non-gaussian features are trivially (almost unavoidably) generated in complex multi-field inflation models~\cite{BartoloReview,Bartolo1,Bartolo2}, or those models with non-canonical kinetic terms~\cite{BartoloReview,Bartolo1,Bartolo2}. They are also generated in alternative to inflation scenarios~\cite{Planck18NG}. However, even in standard single field slow-roll inflation, some tiny amount of non-gaussianity is expected~\cite{Acquaviva,Maldacena}. This small signal consists of two parts: the squeezed part that is proportional to the initial power spectrum tilt $n_s-1$, and the equilateral configuration proportional to the first slow-roll parameter $\epsilon$. This level of non-gausianity was surprising, as before the work by Ref.~\cite{Gangui} the expectation was that canonical single-field non-gaussianities should be $O(\epsilon^2)$. This was expected if the non-gaussianity was originating from the self-coupling of the inflaton field as a $\phi^3$ term\cite{Falk}. However, later on~\cite{Gangui}, it was realized that the non-linear nature of Einstein's equations lead to non-gausianities proportional to the $\epsilon$ term. 

The squeezed term (where one of the momenta goes to zero) can be understood as the effect of the  conserved curvature perturbation $\zeta$ as it re-enters the horizon; it is the time delay for re-entering that causes this long and short-mode coupling. There has been, however, a recent debate on whether the non-gaussian features from canonical single-field inflation can actually be observed. The argument centers around the squeezed limit and the fact that for a (fairly) local observer  the curvature $\zeta$ will be constant, and thus can be trivially removed by an appropriate coordinate transformation. However, these arguments only apply to (strictly) local observations; access to distant observers does imply that the local frame does not see a constant curvature any longer, thus the equivalence principle cannot be used (see e.g. Ref~\cite{NGgauge} and references therein for more on this debate).

In a series of recent works~\cite{qf1,qf2}, we have put forward a new physical picture of the early Universe, assuming an early quasi de-Sitter phase. The key idea lies in representing the power spectrum of the primordial fluctuations of the curvature in terms of a cosmological quantum Fisher information function. In this paper we want to use this picture to describe the expected amount of non-gaussianity from the early universe. 

Our main findings are: the early universe has a minimum level of non-gaussianity, that expressed in terms of the parameter $f_{\rm NL}$ is $\sim 0.03$. We also show that there are contributions from both the equilateral and the squeezed shapes. Finally, it is shown that the squeezed limit is a physical non-gaussianity and not a gauge mode, as otherwise the quantum Fisher would be cero, which is unphysical; it would imply that time and energy can be measured with zero uncertainty simultaneously. Only in the exact limit of momentum equal zero, can this mode be re-absorbed as a global curvature. While the predicted value of non-gaussianity is small ($f_{\rm NL} =0.03$), it could be reached by upcoming full sky galaxy surveys~\cite{Carbone,Spherex}. Finally, we point out to a new pure quantum effect: {\it clock bias}, this induces a very small, but measurable, tilt in the initial power spectrum ($\sim 10^{-4}$). This effect could provide valuable insights into the quantum nature of the early Universe.

\section{Quantum estimation theory: a General Relativity correspondence.}

In this section, we will try to establish a potential correspondence between General Relativity (GR)  and quantum estimation theory for the particular case of inflationary cosmology. From the point of view of GR, we deal with a Lagrangian where we have a scalar field $\phi$, the inflaton, with a given potential $V(\phi)$, coupled to the Einstein-Hilbert action. We are interested in discovering the effective Lagrangian for small quantum fluctuations i.e. for fluctuations $\delta \phi$ of the inflaton field as well as the fluctuations of the metric around a classical solution corresponding to the quasi de Sitter evolution driven by the slow rolling of the scalar field. Re-parametrization invariance reduces the number of independent scalar degrees of freedom to one (Galilean invariance). We can use the ADM formalism, where we introduce a foliation of the space time in hyper-surfaces $\Sigma(t)$ defining a Hamiltonian system with coordinates $h_{i,j}$ and $\phi$ and the corresponding canonical conjugated momenta. The corresponding Hamiltonian is a constrained system with two constraints corresponding to total energy and total momentum equal to zero. The Lagrange multipliers defining these constraints are the lapse function $N$ and the shift functions $N_i$ respectively. Solving the constraints on the quasi de Sitter solution allows us to find the effective Lagrangian that contains both the information about power spectrum as well as non gaussianities. 

Let us now define the quantum mechanical analog as follows. We use a quantum density matrix $\rho(t)$ and a time dependent Hamiltonian $L_t$ given by the equation
\begin{equation}\label{equ}
\frac{d\rho}{dt}= \frac{L_t\rho(t)+\rho(t)L_t}{2}
\end{equation}
The Fisher information function for this system is 
\begin{equation}\label{Fisher}
F(t) =Tr[\rho(t) L_t^2]
\end{equation}
In particular, this means that 
\begin{equation}\label{constraint}
tr[\rho(t)L_t]=0
\end{equation}
for any time $t$. At this point the first correspondence is
\begin{equation}
tr[\rho(t)L_t]=0 \Leftrightarrow H=0
\end{equation}
i.e. the GR constraint in the ADM formalism corresponding to zero total energy ($H$ is the Hubble parameter). In GR this is the constraint we use to extract the lapse function $N$. In order to introduce the input about the quasi de Sitter evolution what we do is to estimate the Fisher function as the second derivative of the relative entanglement entropy. This was done in Ref.~\cite{qf1} leading to
\begin{equation}\label{Fisher2}
F(t) \sim \epsilon(t)^2
\end{equation}
for $\epsilon$ the slow-role parameter. Once we use this input, the former correspondence implies that the GR constraint $H=0$ becomes equivalent to impose 
(\ref{Fisher}) and (\ref{constraint}). 

The next step is to define the clock operator as the quantum estimator of time~\cite{Paris}. This operator is defined by
\begin{equation}\label{clock}
\hat T(t)= t \mathbb{I} +\frac{L_t}{F(t)},
\end{equation}
where $\mathbb{I}$ is the identity operator. The clock operator defined in (\ref{clock}) is the maximally efficient quantum estimator for time i.e. it saturates the Cramer-Rao bound for its variance. This operator contains two pieces. The component $t \mathbb{I}$  is independent of the dynamics, while the non trivial quantum part $\frac{L_t}{F(t)}$  depends on the time evolution of the underlying quantum state $\rho(t)$ and the exact quantum Fisher function $F(t)$. Thus this clock operator is self consistently defined with the dynamics.

Since we are interested in small fluctuations, we will define the operator 
\begin{equation}
\hat \delta(t) = \hat T(t)-t \mathbb{I}
\end{equation}
In inflation, the natural clock is defined by the classical rolling of the inflaton field. This leads us to define a new operator
\begin{equation}\label{zeta}
\hat \zeta(t) = \dot \phi \hat \delta(t) \frac{1}{M_P}
\end{equation}
This defines our second correspondence, namely
\begin{center}
$\hat \zeta(t)$ $\Leftrightarrow$ scalar curvature fluctuations
\end{center}
The deep geometrical meaning of this correspondence will become clear in a moment. 

Quantum estimation theory allows us to evaluate $\langle \hat \zeta \hat\zeta \rangle$ as the corresponding variance, namely
\begin{equation}\label{delta}
\langle \hat \zeta \hat\zeta \rangle = \frac{\dot \phi ^2}{F(t)M_P^2} = \frac{\dot \phi ^2}{M_P^2} \Delta^2(\hat \delta(t))\equiv \Delta^2(t)
\end{equation}
leading to the correspondence:
\begin{center}
scalar power spectrum ( at horizon exit ) $\Leftrightarrow$ quantum variance of $\hat \delta(t)$
\end{center}
Note that we are taking the value of this variance saturating the Cramer-Rao inequality.

We can define a proper time operator as
\begin{equation}
\hat \tau(t) = \frac{\hat \zeta}{H}
\end{equation}
The corresponding proper time Fisher function is simply given by
\begin{equation}\label{proper}
\tilde F (t) \sim \epsilon
\end{equation}
The first order quantum contribution to the lapse function $N$ that we shall denote  $N_1$ i.e. $N=1+N_1$ is then determined as usual by
\begin{equation}
N_1 \sim \frac{d\tau}{dt}
\end{equation}
The final step is to understand the geometrical meaning of this lapse function from quantum estimation theory. We can define the quantum metric $g^q_{00}$ by
\begin{equation}
d\tau = \sqrt{g^q_{00}} dt
\end{equation}
that leads at first order to
\begin{equation}
g^q_{00} \sim G^F_{00}
\end{equation}
with $G^F_{00}$ the Fisher metric
\begin{equation}
G^F_{00} = \tilde F(t)
\end{equation}
leading to the final correspondence:

\begin{center}
quantum correction to $g_{00}$ $\Leftrightarrow$ quantum Fisher metric for the time estimator.
\end{center}
Finally we can define a {\it quantum tilt} for the quantum Fisher function as 
\begin{equation}
-\frac{d ln (F)}{dt}
\end{equation}
that for the cosmological Fisher function (\ref{Fisher}) leads to
\begin{equation}
-\frac{d ln (F)}{dt} = -2 H \eta
\end{equation}
for $\eta= \frac{\dot \epsilon}{\epsilon H}$ the slow roll parameter.

Let us finish this section with two general comments. First of all, note that the real quantum parameter is $\epsilon$. Indeed, both the quantum Fisher function as well as the operator $L_t$ go to zero in the limit $\epsilon =0$. From the quantum point of view $\epsilon=0$ implies that the distance between the different states of the Universe along its primordial evolution is zero or, in other words, that there is no relative entanglement entropy generation. The second comment refers to the key role of the quantum time uncertainty relation in primordial cosmology. In essence, the whole point of the primordial quantum fluctuations lies in understanding the proper time fluctuations on a hyper-surface of constant energy driven by quantum time energy uncertainty. What we are just presenting lies in understanding this quantum uncertainty from the point of view of quantum estimation theory. Recall that there is not such a thing in standard quantum mechanics as a time operator and that the time energy uncertainty should be understood statistically. 

\section{Primordial Gaussianity}
 It is easy to check that
\begin{equation}
Tr(\rho(t) \hat \delta(t)^{2n+1}) =0
\end{equation}
Using our table of correspondences from the previous section, this primordial gaussianity appears as a consequence of the constraint $H=0$, determined by re-parameterization invariance. 
The gaussian distribution is completely determined by the variance
\begin{equation}
\Delta^2(\hat \delta(t)) = Tr(\rho(t) \hat \delta(t)^2) 
\end{equation}
which as already discussed is determined by the quantum Fisher $F(t)$.

Using the operators $\hat \zeta(t)$ defined in (\ref{zeta}), the power spectrum is given by $\Delta^2(t)$ defined in (\ref{delta}). The tilt is now defined by
\begin{equation}\label{tilt}
\frac{d \Delta^2(t)}{dt} = \alpha \Delta^2(t)
\end{equation}
In order to understand this definition let us recall the standard definition of the tilt. In the standard approach we define the Fourier components $\hat \zeta_k(t)$ with
\begin{equation}
\langle \hat\zeta_k(t) \hat\zeta_{-k}(t) \rangle = \frac{1}{k^3} \Delta^2(t)
\end{equation}
with $\Delta^2(t) = \frac{H^2(t)}{\epsilon(t)}$ for $t$ determined by the horizon exit condition
\begin{equation}
k=aH
\end{equation}
In these conditions the tilt $(n_s-1)$ is defined as
\begin{equation}\label{one}
\frac{dln(\Delta^2(t))}{dlnk} = (n_s-1)
\end{equation}
Now using $k=aH$ you can transform (\ref{one}) into
\begin{equation}
\frac{d\Delta^2(t)}{dt} = H (n_s-1) \Delta^2(t)
\end{equation}
That is our equation (\ref{tilt}) for $\alpha=H (n_s-1)$ that leads to $n_s-1 = -2\epsilon-\eta$.

In our simple approach we are working with $\hat \zeta(t)$ and not with the Fourier components $\hat \zeta_{k}(t)$. However whenever we are interested in the physical observables evaluated at horizon exit we can define a map
\begin{equation}\label{map}
 \hat\zeta_k(t) \rightarrow \hat \zeta(t(k))
\end{equation} 
for $t(k)$ defined by the horizon exit condition
$k =a(t(k))H(t(k))$.

As usual we can define the running of $n_s$ by
\begin{equation}\label{run}
\alpha = \frac{d ln(n_s)}{Hdt}
\end{equation}
that is given by $\alpha = - 2\eta \epsilon - \frac{\dot \eta}{H}$ which is bounded to be smaller than $0.01$.

\section{Primordial Non Gaussianities}
Let us start with
\begin{equation}
\langle \hat \zeta(t)^3 \rangle = Tr[\rho(t) \hat \zeta(t)^3],
\end{equation}
which is zero. The sort of correlations in which we are interested are 
\begin{equation}
\langle \hat \zeta_{k_1}(t) \hat \zeta_{k_2}(t) \hat\zeta_{k_3}(t) \rangle
\end{equation}
with $k_1+k_2+k_3=0$. Let us now consider the corresponding correlation defined by the map (\ref{map}) i.e.
\begin{equation}
\langle \hat \zeta (t(k_1)) \hat \zeta(t(k_2)) \hat \zeta(t(k_3))\rangle
\end{equation}
Let us now impose the conservation of momentum in two limits. The squeezed limit where $k_1$ is assumed to be very small and the equilateral limit where we consider that the three $k's$ are of the same order i.e. $t(k_1)\sim t(k_2) \sim t(k_3)$.

In the squeezed limit we can represent the amplitude in the corresponding quantum mechanical approach as

\begin{equation}
Tr(\rho(t(k_2)) \hat \zeta(t(k_1)) \hat \zeta^2(t(k_2)+\delta)) 
\end{equation}
where $\delta$ is determined by $k_1$. Since $k_1$ is very small, we can approximate the former expression using a formal cluster decomposition as:
\begin{equation}
\zeta(t(k_1)) \langle \hat \zeta(t(k_2)+\delta) \hat \zeta(t(k_2)+\delta) \rangle =\zeta(t(k_1))\Delta^2(t+\delta)
\end{equation}
where we distinguish between the classical $\zeta(t)$ and the quantum $\hat \zeta(t)$.

Using (\ref{tilt}) we get
\begin{equation}
\alpha \zeta(t(k_1))  \delta \Delta^2(t)
\end{equation}
where now both $\zeta(t(k_1))$ and $\delta$ are c-numbers both determined by $k_1$. 
This can be easily evaluated in the squeezed limit
\begin{equation} \label{time}
 \frac{\zeta(t(k_1))}{H(t(k_1))} \sim \delta 
\end{equation}
  with $\zeta(t(k_1))$ equal to the minimal statistical  uncertainty $\sqrt{\Delta^2(t(k_1))}$. The result is
\begin{equation}
\langle \hat \zeta(t(k_1)) \hat \zeta^2(t(k_2)+\delta)\rangle = (n_s-1) \Delta^2(t(k_1)) \Delta^2(t(k_2))
\end{equation}
with $(n_s-1)$ evaluated at the time $t(k_2)$. Reintroducing the explicit momentum dependence, we get the well known result
\begin{equation}
\langle \hat \zeta_{k_1} \hat \zeta_{k_2}\hat \zeta _{k_3} \rangle \sim \frac{1}{k_1^3 k_2^3}(n_s-1) \Delta^2(t(k_1)) \Delta^2(t(k_2))
\label{eq:ns}
\end{equation}

which is the expected non gaussianity in the squeezed limit associated with the tilt \cite{Maldacena}. A question that has triggered some discussion recently (see~\cite{NGgauge} and references therein) is if this form of non gaussianity is really physical. The way to address this question from the former discussion reduces to ask ourselves if in the very soft limit $k_1\sim 0$
the variance $\Delta^2(t(k_1))$ is still different from zero. This will be indeed the case if ${\rm lim}_{k=0} \hat T (t(k)) = t(k)\mathbb{I}$ i.e. if in the $k=0$ limit the quantum time estimator is just the classical time with zero variance. This question is subtle and requires to take into account the asymptotic soft limits i.e. infrared physics. Very likely, in reality, this formal limit should be constrained by an infrared cutoff $K > 0$ with $\frac{1}{K}$ of the order of the size of the visible Universe. In this case this variance cannot be set to zero and the squeezed non gaussianity remains physical for $k_1 \geq K$. 

\subsection{Equilateral non gaussianity}
Let us now consider the case where the three momenta are of the same order $k$. In this case, if we insert the three $\hat \zeta(t)$ operators at the same time $t(k)$, conservation of momentum implies that the non vanishing amplitude is

\begin{equation}
Tr(\rho(t(k)+\delta) \hat \zeta(t(k))^3)
\end{equation}
with, see (\ref{time}),
\begin{equation}\label{delta2}
\delta \sim \frac{\zeta(t(k))}{H(t(k))} 
\end{equation}

Now we can expand $\rho(t+\delta)$ using its equation of motion given by (\ref{equ}). We get

\begin{eqnarray}
Tr(\rho(t(k)+\delta) \hat \zeta(t(k))^3)  & =& \\ \nonumber
 \delta \frac{H(t(k)F(t(k))}{\dot \phi(t(k)} Tr(\rho(t(k)) \hat \zeta(t(k))^4) & = & \\ \nonumber
 \delta \frac{H(t(k))F(t(k))}{\dot \phi} \Delta^4(t(k))
\end{eqnarray}

In this case the non gaussianity is given not by the tilt but by $ \delta \frac{F(t)H(t(k))}{\dot \phi}$. Using (\ref{delta2}) we get for $|k_1|\sim |k_2|\sim |k_3| \sim |k|$
\begin{equation}
\langle \hat \zeta_{k_1}\hat \zeta_{k_2}\hat\zeta_{k_3}\rangle
\sim \epsilon(t(k))\Delta^4(t(k))\frac{1}{k^6}
\end{equation}
in agreement with the results of \cite{Maldacena}. This implies

\begin{equation}
{\rm non-gaussianity}  \sim 4\pi \epsilon(t(k))
\end{equation}
where we have reintroduced the numerical factors and used the cosmological Fisher function (\ref{Fisher2}).

It is important to stress the physical origin of the generic non gaussianity in our approach. The fundamental reason is simply that the operator $L_t$ defining the time evolution of the quantum state represented by $\rho(t)$ is itself part of the operator $\hat \zeta$. This is the reason the time evolution of $\rho$ generates for $\epsilon\neq 0$ primordial non gaussianity. Note also that this non gaussianity is very much model independent and is completely determined by the value of $\epsilon$. 

Using the standard parametrization~\cite{VerdeWang,Komatsu}
\begin{equation}
\zeta = \zeta_g + \frac{3}{5} f_{NL} \zeta_g^2
\end{equation}
 we get, using the CMB constraint discussed in \cite{qf2}
 \begin{equation}
 f_{NL} \sim 12 \pi \epsilon \sim 0.03
\end{equation}

For the squeezed configuration we obtain that $f_{\rm NL} \propto (n_s -1)$ as per~(\ref{eq:ns}). As discussed above, this term can be only suppressed in the limit $k=0$ where $\hat T (t(k)) = t(k)\mathbb{I}$. If we set as a maximal physical scale the size of the visible Universe  i.e. an infrared cutoff with $K>0$ this effect cannot be suppressed  for an observer with some information beyond the local patch. A different question is if the squeezed non gaussianity will induce some bias for the power spectrum. If we introduce an infrared cut-off corresponding to the co-moving size $R$ of the local patch, we observe that the local re-parametrization induced by the long wave mode with $k\sim 0$ can be compensated by the corresponding change of $R$ provided the variation of the long wave mode in a region of co-moving size $R$ is negligible. However, any bias that is not strictly local will see this squeezed primordial non-gaussianity.

\section{ A possible primordial bias signature: {\it clock bias}.}

The notion of bias in cosmology was introduced as a way to parametrize our ignorance about structure formation dynamics~\cite{Kaiser}. The typical case of bias, the galaxy bias, defined by
\begin{equation}
\delta_g =b \delta_m
\end{equation}
relates the fluctuations of a matter tracer as can be the galaxy or halo distribution $\delta_g$ to the fluctuation of the matter density. This bias parametrizes our ignorance about galaxy ( or any other tracer ) formation. A bias of this type leads for the two point correlation function to the relation
\begin{equation}\label{bias}
\Delta^2_g= b^2 \Delta^2_m
\end{equation}
The bias parameter can obviously depend on time and on scale. An almost trivial example of bias can be introduced for the primordial spectrum itself. Namely we can interpret $\Delta^2 = \frac{H^2}{M_P^2} \frac{1}{\epsilon}$ as  
(\ref{bias}) with
\begin{equation}
b^2 \equiv \frac{H^2}{M_P^2}
\end{equation}
relating the primordial spectrum for curvature fluctuations to a formal primordial power spectrum $\frac{1}{\epsilon}$ that we can interpret as the power spectrum for {\it proper time fluctuations} ( see (\ref{proper})
\begin{equation}
\Delta^2 = \frac{H^2}{M_P^2} \frac{1}{\epsilon} = b^2 \Delta^2_{t}
\end{equation}
In other words $\frac{H^2}{M_P^2}$ can be interpreted as {\it the bias factor
for curvature fluctuations as a tracer of time fluctuations}. If we use the Gibbons-Hawking relation
$\frac{H^2}{M_P^2}= \frac{1}{N}$ we observe that this bias is equal to one for $H=M_P$ and diminishes during the inflationary period. 
In this case curvature fluctuations play the role of {\it tracers} of real {\it proper time fluctuations}.

The former discussion is purely classical, so the natural question is: Can we define a notion of {\it quantum bias} ? To address this question, we will consider just the time estimator defined in (\ref{clock}).

This operator is by construction unbiased i.e.
\begin{equation}
\langle \hat T(t)\rangle - t =0,
\end{equation}
as well as it is the maximally efficient estimator in the sense that it saturates for its variance the Cramer-Rao bound. It is convenient to interpret the two pieces of $\hat T$ as classical $t$ and quantum $\frac{L_t}{F(t)}$.

Now, a natural definition of {\it quantum bias} for this time estimator can be introduced using in the definition of the quantum estimator an operator $\tilde L_t$ that is not the exact operator $L_t$ associated with the time evolution of the underlying quantum state. An obvious case is when we lack some information on the dynamics and we define our estimator on the basis of a reduced dynamical information. In this case the so defined operator can be biased. Let us in what follows consider a very simple example where
the  {\it bias clock} is defined as a clock operator where we set the $t$ entering into the quantum component to be some fixed value $\tilde t$:
\begin{equation}
\hat T_b(t) = t \mathbb{I} + \frac{L_{\tilde t}}{F(\tilde t)}
\end{equation}
In this sense the clock operator for $t=\tilde t$ agrees with the unbiased clock but the bias is appearing in the subsequent time evolution. Indeed, the classical clock continues working but the quantum part is frozen by construction to the value $\tilde t$ used to define the bias. 

The bias induced by this definition is simply given by
\begin{equation}
\langle \hat T_b(t) \rangle - t = Tr(\phi(t) \frac{L_{\tilde t}}{F(\tilde t)}) \equiv b_q
\end{equation}
We can now compute the bias for the operator $\hat \zeta_b(t) = \dot \phi \delta_b(t) \frac{1}{M_P}$ for $\delta_b(t) =\hat T_b(t)-t \mathbb{I}$  created after a time $\delta(t)$ starting at time $t=\tilde t$. Using $\zeta= \delta(t)H$ i.e assuming that the bias is created only before the time at which the mode is frozen--the horizon exit--the bias is given by
\begin{equation}
b_q= \sim \frac{\zeta}{H}
\end{equation}
The effect of this bias on the power spectrum is easy to evaluate using the modification of the quantum Cramer-Rao due to bias, namely
\begin{equation}
\Delta_b^2(t) = (1+\dot b_q)^2 \Delta^2(t)
\end{equation}
that leads to
\begin{equation}
\Delta^2_b(t) \sim (1 + N_1) \Delta^2(t)
\end{equation}
for $N_1$ the quantum correction to the lapse function i.e. $N=1+N_1$. Thus, we conclude that there is a potential  primordial bias of order $N_1$ whenever we use a fixed frozen time for the definition of the time quantum estimator. 

This bias corrects the value of the power spectrum spectral index in a way that is small, namely
\begin{equation}
(n_s)_b - n_s \sim \frac{\sqrt{\epsilon} \eta}{2} \sim 0.001
\end{equation}
The time dependence of this bias also modifies the value of $\alpha$, the running of the spectral index, defined in (\ref{run}) in again a small quantity
\begin{equation}
\alpha_b - \alpha \sim \eta^2 \sqrt{\epsilon} \sim 10^{-4}
\label{eq:cbias}
\end{equation} 

However, these corrections are in principle not negligible as they can actually be measured. The predicted bias in~\ref{eq:cbias} is only a factor few smaller than current limits by Planck18~\cite{Planck18} on the running of the spectral index of the primordial power spectrum. There are two possible avenues to measure experimentally this clock bias: the first one is by comparing the spectral tilt and the running of the spectral index in different patches of the sky; these patches should have a clock bias of the order of what we have computed in eq.~\ref{eq:cbias} between each order, thus our prediction is that the running should be different among patches by $\sim 10^{-4}$. The second route is to measure how time changes with redshift, like, e.g., using redshift drift on our local frame by upcoming experiments~\cite{Racca}, as this will manifest in a bias by setting our clock to $H_0$.  If we use a clock in our Earth laboratory and compare it with a clock at cosmological distances (the quantum clock), we should observe a deviation of how this clock changes with respect to the classical expectation, i.e., a $\Lambda$CDM model.

It is important to stress that this bias is associated with the potential lack of {\it synchronization} between the classical and the quantum time correction contributions to the clock operator. Non gaussianities were derived using a similar effect but where the de-synchronization in three point functions is due to the momentum conservation for the Fourier components in the horizon exit prescription. This potential primordial bias is in origin very different from other forms of bias associated with squeezed non gaussianity where what is considered is space re-parametrizations induced by the large (small momentum ) component of $\zeta$. In those cases the bias can be locally, but not globally, reabsorbed by a compensating space diffeomorphism in the local patch. In the present case we are considering instead {\it clock bias}. A measurement of this {\it clock bias} could provide precious insight into early universe physics.

\section{Summary}

We have described the early Universe using the quantum information picture by describing it as a quantum density matrix with the corresponding quantum Fisher. In our picture the primordial power spectrum of  density perturbations is the inverse of this Fisher matrix. 

We have computed the higher order correlation functions and found that the early universe has very small non-gaussianites. These are in the equilateral ($f_{\rm NL} \sim 0.03$) and squeezed shape ($f_{\rm NL} \sim n_s-1$). The physical origin of the equilateral shape is the non-linearity of Einstein's equations i.e. the evolution of the underlying quantum state.  For the squeezed shape we have shown that this effect is real and could only be surpresed if quantum effects were zero in the early universe, i.e. if the quantum clock was the same as the classical clock. If we set as a maximal physical scale the size of the visible Universe, this effect cannot be suppressed for an observer with some information beyond the local patch.

We have identified a new effect {\it clock bias}, which is a pure quantum effect and introduces a bias in the spectral tilt and running of the power spectrum of order $\sim 10^{-4}$. This predictions of our model complement previous ones~\cite{qf2} where we have shown that the expected level of primordial gravitational waves produces a tensor-to scalar ratio of $r = 0.01$. These features could be measured in the current decade with ground based Stage-IV CMB polarization experiments and with full-sky (optical or 21cm HI) galaxy surveys. An experimental detection would  open up the window to understand our Universe as a quantum physical system.

\acknowledgments

 The work of CG was supported by grants SEV-2016-0597, FPA2015-65480-P and PGC2018-095976-B-C21. Funding for this work (RJ) was partially provided by the Spanish science ministry under project PGC2018-098866-B-I00.

\end{document}